\documentclass[12pt]{article}

\newcommand{\be}{\begin{equation}}
\newcommand{\ee}{\end{equation}}
\newcommand{\ba}{\begin{eqnarray}}
\newcommand{\ea}{\end{eqnarray}}

\newcommand{\bc}{\begin{center}}
\newcommand{\ec}{\end{center}}
\newcommand{\bo}{\begin{order}}
\newcommand{\eo}{\end{order}}
\newcommand{\bi}{\begin{itemize}}
\newcommand{\ei}{\end{itemize}}
\newcommand{\bp}{\begin{picture}}
\newcommand{\ep}{\end{picture}}
\newcommand{\beq}{\begin{equation}}
\newcommand{\eeq}{\end{equation}}
\newcommand{\bea}{\begin{eqnarray}}
\newcommand{\eea}{\end{eqnarray}}
\newcommand{\bfi}{\begin{figure}[tbp] \unitlength=1cm}
\newcommand{\efi}{\end{figure}}
\newcommand{\bfc}{\begin{figure}[tbp] \unitlength=1cm%
         \begin{center}}
\newcommand{\ecf}{\end{center} \end{figure}}
\renewcommand{\bar}{\overline}
\newcommand{\derpre}{\stackrel{.}}
\newcommand{\dersec}{\stackrel{..}}

\newlength\cadredim
\begin{document}
\setlength{\baselineskip}{.5cm}
\renewcommand{\thefootnote}{\fnsymbol{footnote}}
\newcommand{\lp}{\left(}
\newcommand{\rp}{\right)}

\begin{center}
\centering{{\bf \LARGE Solid friction at high sliding velocities\,:\\ an
explicit 3D dynamical
SPH approach}
\vskip 1cm
C. Maveyraud $^1$, W. Benz $^2$, G. Ouillon $^1$, A. Sornette $^1$ and D.
Sornette $^{1,3}$\\
\vskip 0.5cm
$^1$ LPMC, CNRS UMR6622 and  Universit\'e de Nice-Sophia Antipolis\\
B.P. 71, 06108 NICE Cedex 2, France\\
$^2$ Physikalisches Institut, Universitaet Bern\\
Sidlerstrasse 5, CH-3012 Bern, Switzerland\\
$^3$ IGPP and ESS department, UCLA, Box 951567\\ Los Angeles, CA
90095-1567, USA
}
\end{center}

\vskip 1cm
{\bf Abstract}: We present realistic 3D numerical simulations of elastic 
bodies sliding on top of each other in a regime of velocities ranging 
from meters to tens of meters per second using the so-called Smoothed 
Particle Hydrodynamics (SPH) method. This allows us to probe intimately 
the response of the bodies and the nature of the friction between them.
Our investigations are restricted to regimes of pressure and roughness 
where only elastic deformations occur between asperities at the contact
surface between the slider block and the substrate. In this regime,
solid friction is due to the generation of vibrational radiations which
are subsequently escaping to infinity or damped out in which case energy 
is dissipated. We study periodic commensurate and incommensurate asperities 
and various types of disordered surfaces. In the elastic regime 
studied here, we report the evidence of a transition from zero 
(or non-measurable $\mu < 0.001$) friction to a finite friction as the normal 
pressure increases above about $10^6~Pa$.  For larger normal pressures 
(up to $10^9~Pa$), we find a remarkably universal value for the friction 
coefficient $\mu \approx 0.06$, which is independent of the internal 
dissipation strength over three order of magnitudes, and independent
of the detailled nature of the slider block-substrate interactions. We 
find that disorder may either decrease or increase $\mu$ due to the 
competition between two effects: disorder detunes the coherent vibrations 
of the asperties that occur in the periodic case, leading to weaker 
acoustic radiation and thus weaker damping. On the other hand, large
disorder leads to stronger vibration amplitudes at local asperities and 
thus stronger damping.  Our simulations have confirmed the existence of 
jumps over steps or asperities of the slider blocks occurring at 
the largest velocities studied ($10~m/s$). These jumps lead to chaotic
motions similar to the bouncing-ball problem. We find a velocity
strengthening with a doubling of the friction coefficient as the 
velocity increases from $1~m/s$ to $10~m/s$. This reflects the increasing 
strength of vibrational damping.

\pagebreak

\section{Introduction}

Solid friction has a long scientific history, starting probably in 
the western world about 500 years ago with the work of Leonardo de Vinci, 
continuing with the empirical Amontons'laws two centuries later and 
Coulomb's investigations of the influence of slipping velocity on friction 
in the XVIII century. Only three decades ago was it recognized that friction 
plays probably a fundamental role in the mechanics of earthquakes [{\it 
Brace and Byerlee}, 1966]. Rock mechanicians consider an earthquake as a 
stick-slip event controlled by the friction properties of the fault, i.e. 
the destabilization of a weak part of the crust.  This formulation has been 
shaped by laboratory experiments performed under a variety of pressure and 
temperature conditions (which however reproduce only imperfectly the 
conditions prevailing in the crust).  Numerous laboratory experiments have 
been carried out to identify the parameters that control solid friction and 
its stick-slip behavior [{\it Persson and Tosatti}, 1996]. The most 
significant variables appear to be the mineralogy, the porosity, the 
thickness of the gouge, the effective pressure, the temperature and the 
water content [{\it Byerlee et al}, 1968; {\it Brace}, 1972; {\it Beeman 
et al.}, 1988; {\it Gu and Wong}, 1991; {\it Johansen et al.}, 1993; 
{\it Streit}, 1997].  Low velocity experiments have established that solid 
friction is a function of both the velocity of sliding and of one or several 
state parameters, roughly quantifying the true surface of contact [{\it Brace}, 
1972; {\it Dieterich}, 1972; 1978; 1979; 1992; {\it Ruina}, 1983; {\it Cox}, 
1990; {\it Beeler et al.}, 1994; 1996; {\it Baumberger and Gauthier}, 1996; 
{\it Scholz}, 1998].

The Ruina-Dieterich laws constitute the basic ingredients in most models
and numerical elastodynamic calculations that attempt to understand the 
characteristics of earthquake sources. A recent lively debate has been whether
space-time complexity in earthquake sequences can occur on an homogeneous
fault solely from the nonlinear dynamics [{\it Shaw}, 1993; 1995; 1997; {\it 
Cochard and Madariaga}, 1994; 1996] associated with the slip and the velocity 
dependent friction law [{\it Dieterich and Kilgore}, 1994; {\it Dieterich}, 
1992], or, does it necessarily require the presence of quenched heterogeneity 
[{\it Rice}, 1993; {\it Benzion and Rice}, 1993; 1995; {\it Knopoff}, 1996]?
It is now understood that complexity can emerge purely from the nonlinear
laws but heterogeneity is probably the most important factor dominating the 
multi-scale complex nature of earthquakes and faulting [{\it Ouillon et al.}, 
1996]. It is also known to control the appearance of self-organized critical 
behavior in a class of models relevant to the crust [{\it Sornette et al.}, 
1995; {\it Shnirman and Blanter}, 1998].

A well-known and serious limitation of these calculations based on laboratory 
friction experiments is that the friction laws have been determined using sliding 
velocities no more than about $1$~cm/s, i.e.  orders of magnitude below the 
sliding velocity of meters or tens of meters believed to occur during earthquakes. 
 The validity of extrapolations, especially the velocity weakening dependence,
has yet to be demonstrated. This is all the more relevant when one examines 
the underlying physical mechanisms giving rise to the friction laws. At 
low velocity, hysteretic elastic and plastic deformations at the length scale 
corresponding to asperities seem to play a dominant role [{\it Bowden and Tabor}, 
1954; {\it Sokoloff}, 1984; {\it Jensen et al.}, 1993; {\it Dieterich and Kilgore}, 
1994; {\it Caroli and Nozi\`eres}, 1996; {\it Tanguy and Nozi\`eres}, 1996; {\it 
Tanguy and Roux}, 1997; {\it Caroli and Velicky}, 1997; {\it Bocquet and Jensen}, 
1997].

At larger velocities, different mechanisms come into play.  Collisions between 
asperities and transfer of momentum between the directions parallel and perpendicular 
to the motion may become an important mechanism [{\it Lomnitz}, 1991; {\it Pisarenko 
and Mora}, 1994]. This regime is probably relevant to explain the apparent 
low heat flow and thus low friction coefficient observed along the San Andreas 
fault, the so-called heat flow paradox.  This paradox [{\it Bullard}, 1954] 
comes about because, in order to allow for large earthquakes, a fault should have 
a large friction coefficient so that it can store large amount of elastic energy.
However, repeated earthquakes occurring with a large coefficient of dynamical 
friction should give rise to a significant heat flow at the surface which
has not been observed [{\it Henyey and Wasserburg}, 1971; {\it Lachenbruch and 
Sass}, 1980]. One explanation for this low heat flow is that the coefficient of 
friction is low as a result of dynamical effects affecting the friction during the 
earthquake.  Other possibilities, less explored in the literature, involve fluids 
[{\it National Research Council}, 1990] or acoustic vibrations [{\it Melosh}, 1996].  
Several simplified models have recently been explored as possible mechanisms for 
the generation of a low friction. These mechanisms include crack opening modes of 
slip [{\it Brune et al.}, 1993], dynamical collision effects [{\it Lomnitz}, 
1991; {\it Pisarenko and Mora}, 1994], frictional properties of a granular gouge 
model under large slip [{\it Scott}, 1996], space filling bearings with compatible 
kinematic rotations [{\it Herrmann et al.}, 1990], hierarchical scaling [{\it 
Schmittbuhl et al.}, 1996].

Recently, Tsutsumi and Shimamoto [1996; 1997] have reached a completely novel
regime, by performing friction measurements on rotation cylindrical samples at 
velocities up to $1.8$~m/s and for slips of several tens of meters. While 
these results due to several experimental problems are not completely straight-forward
to interpret, they seem to indicate the existence of a change of regime from 
velocity weakening to velocity strengthening and then again to velocity weakening 
at the largest velocities. This last regime seems to be associated to the melting 
of a very thin layer.

The work we report here has been motivated by considerations that different
physical mechanisms might lead to a change of regime in the velocity and slip
dependence of the solid friction law. Thus, one needs to explore the 
high velocity regime in as large a variety of conditions as has been done for the 
low velocity regime. For this purpose, we have developed realistic 3D numerical 
simulations of elastic bodies sliding on top of each other in a regime of 
velocities ranging from meters to tens of meters per second. In this way, we
probe more intimately than any exprimental setup could do the response of the 
bodies and the nature of the friction.

We begin by a short presentation of the SPH method and its implementation in
our context. We follow by a presentation and a discussion of the results 
obtained for surfaces with periodic asperities and then with various types of 
random asperities and conclude.

\section{The numerical model}

\subsection{The SPH method}

We have adapted the Lagrangian method called ``smoothed particle hydrodynamics'' 
(SPH), initially introduced by Lucy [1977] for hydrodynamic problems with fast 
dynamics. This method has the avantage of being simple, elegant, easy to implement 
and to extend with a reasonable precision. Its most recent improved version makes 
it well-suited to treat problems with fast moving interfaces. This was our 
initial motivation to use this property at our advantage to tackle the solid 
friction problem in the regime of large slipping velocities.

In SPH the physical problem is discretized on a mesh whose nodes, the ``particles'' 
are moving or adapting in a Lagrangian manner. Each particles carries with it a 
set of field variables. An interpolating kernel allows to reconstruct the field 
variables everywhere by interpolation. Spatial derivatives are obtained from 
the analytical differentiation of the interpolation kernel. SPH has been used in 
a variety of applications, such as gas dynamics [{\it Monaghan and Gingold}, 1983],
fragmentation of gas clouds [{\it Lattanzio et al.}, 1985], radio jets [{\it 
Coleman and Bicknell}, 1985], impacts [{\it Benz et al.}, 1986; {\it Benz and 
Asphaug}, 1994; {\it Benz et al.}, 1994; {\it Asphaug et al.}, 1998], 
quasi-incompressible fluid flows [{\it Monaghan and Humble}, 1993], material 
rupture [{\it Benz and Asphaug}, 1995], dam rupture, ocean wave propagation and
water falls [{\it Monaghan}, 1994; 1996]. For reviews, see [{\it Monaghan},
1988; 1992; {\it Benz}, 1990].

\subsection{Outline of the approach}

Consider one of the physical field $f(\vec{r})$, which is a function of
position $\vec{r}$. It can always be written as 
\beq
f(\vec{r})=\int f(\vec{r}\,')~\delta(\vec{r}-\vec{r}\,')~\vec{dr'}~,
\label{f1}
\eeq
where $\delta$ is the Dirac function. This suggests
approximating $f(\vec{r})$ by a smoothing kernel as
\bea
\langle f(\vec{r})\rangle=\int
f(\vec{r}\,')~W(\vec{r}-\vec{r}\,',h)~\vec{dr'}~,
\label{f2}
\eea
$W(\vec{r}-\vec{r}\,',h)$ is the interpolating kernel and $h$ is the width
of the kernel and thus the smoothing scale. The kernel is continuous,
differentiable and has the following properties:

\beq
\int W(\vec{r}-\vec{r}\,',h) ~\vec{dr'} = 1
\label{f3}
\eeq
and
\beq
\lim_{h \to 0} W(\vec{r}-\vec{r}\,',h) = \delta(\vec{r}-\vec{r}\,')~.
\label{f4}
\eeq
>From (\ref{f3}) and (\ref{f4}),  $\langle f(\vec{r})\rangle
\stackrel{h \to 0}{\longrightarrow}f(\vec{r})$. From now on, we drop the
arrow and
write $\vec{r}$ as $r$, and similarly for the derivatives.
For a fluid of density $\rho(r)$, equation (\ref{f2}) reads
\beq
\langle f(r)\rangle = \int \left[\frac{f(r')}{\rho(r')}\right] W(r-r',h)
\rho (r') dr'~.
\eeq
Discretizing space in $N$ elements of masses $m_j$, the integral becomes
\beq
\langle f(r_{i})\rangle = \sum_{j=1}^{N} m_{j} \frac{f_{j}}{\rho_{j}}
W(r_{i}-r_{j},h)~,
\eeq
with $f_{j} \equiv f(r_{j})$. Replacing $f(r)$ by $\rho(r)$, this yields
the following expression of the fluid density
\beq
\langle \rho(r_{i})\rangle=\sum_{j=1}^{N}m_{j}W(r_{i}-r_{j},h)~.
\eeq
This equation has the following interpretation [{\it Benz}, 1990]\,: each
particle of mass $m_{j}$ is spatially smoothed out according to the 
kernel space dependence, which can be seen as its spatial density 
distribution. The density at any point in space is obtained by summing 
all contributions from all particles at this point. The term Smoothed 
Particle Hydrodynamics expresses this interpretation.

The gradient of $f$ is similarly obtained as
\beq
\langle \nabla f(r)\rangle = \int \nabla f(r') ~W(r-r',h) ~dr'~.
\eeq
Integrating by parts and using the fact that $W$ goes to zero sufficiently
fast so that the surface terms are negligible, one obtains
\beq
\langle \nabla f(r)\rangle = \int \nabla W(r-r',h)~f(r')~dr'~,
\eeq
with $\nabla W(r-r',h)$ being the gradient with respect to $r$.
Discretizing, this yields
\beq
\langle \nabla f(r_{i})\rangle = \sum^{N}_{j=1} \frac{m_{j}}{\rho_{j}}
f_{j} \nabla_{i} W_{ij}~,
\eeq
where $\nabla_{i}$ is the gradient with respect to the coordinate of the
$i$-th particle and
$W_{ij} \equiv W(r_{i}-r_{j},h)$.

Several choices are possible for the kernel, as long as the conditions
(\ref{f3}) and (\ref{f4}) are fulfilled. Kernels constructed on spline
functions have several advantages [{\it Monaghan and Lattanzio}, 1985]\,:
\bea
W(r,h)=\frac{1}{\pi h^{3}} \cdot \left\{
\begin{array}{cc}
1-\frac{3}{2}v^{2}+\frac{3}{4}v^{3}	&\mbox{if\ } 0\leq v\leq 1 \\
\frac{1}{4}(2-v)^{3}			&\mbox{if\ } 1\leq v\leq 2 \\
0					&\mbox{else,}
\end{array}
\right.
\eea
where $v=\frac{r}{h}$. This kernel has compact support, i.e. interactions
between particles vanish for $r>2h$. Only contributions from neighbors 
need to be accounted for instead of $N^{2}$ contributions. The second 
derivative of $W(r,h)$ is continuous and the error in the estimation of 
the interpolation is of order ${\cal{O}}(h^{2})$.  At the beginning of 
the calculation, the values of the kernel and its gradient for different 
values of $v$ are calculated and stored in a array. This allows to 
decrease the calculation time.

\subsection{Equations of motion}

The formulation we use is borrowed from [{\it Stellingwerf and Wingate}, 1992;
{\it Benz and Asphaug}, 1995].
The equation of mass conservation
\beq
\frac{d\rho}{dt} + \rho \frac{\partial v^{\alpha}}{\partial x^{\alpha}}=0~,
\eeq
where $v$ is the local velocity  along $\alpha$, gives after SPH discretization
\beq
\frac{d\rho_{i}}{dt}=\sum_{j=1}^{N}m_{j}(v_{i}^{\alpha}-v_{j}^{\alpha})~\nabla_{
i}^{\alpha}W_{ij}~.
\eeq

The equation of momemtum conservation (neglecting gravity)
\beq
\frac{dv^{\alpha}}{dt}=\frac{1}{\rho}\frac{\partial \sigma^{\alpha
\beta}}{\partial x^{\beta}}~,
\eeq
where $\sigma^{\alpha \beta}$ is the stress tensor defined by
\beq
\sigma^{\alpha \beta}=-P\delta^{\alpha \beta}+S^{\alpha \beta}
\eeq
where $P$ is the pressure, $S^{\alpha \beta}$ is the deviatoric stress
tensor with
zero trace, and $\delta^{\alpha \beta}$ is the Kronecker symbol, becomes
\beq
\frac{dv_{i}^{\alpha}}{dt}=\sum^{N}_{j=1}m_{j}\left(\frac{\sigma_{i}^{\alpha \beta}}{\rho_{i}^{2}}
+\frac{\sigma_{j}^{\alpha
\beta}}{\rho_{j}^{2}}\right)\nabla_{i}^{\beta}W_{ij}~,
\eeq
written in a symmetric form. It is easy to check that the total momemtum is
conserved by multiplying this equation by $m_{i}$ and verifying the exact 
symmetry in $i$ and $j$.

The equation of energy conservation
\beq
\frac{du}{dt}=-\frac{P}{\rho}\frac{\partial v^{\alpha}}{\partial x^{\alpha}}
+\frac{1}{\rho}S^{\alpha \beta}\derpre{\epsilon}^{\alpha \beta}~,
\eeq
where $\derpre{\epsilon}^{\alpha \beta}$ is the tensor of the rate of
deformations defined by
\beq
\derpre{\epsilon}^{\alpha \beta}=\frac{1}{2}\left(\frac{\partial
v^{\alpha}}{\partial x^{\beta}}
+\frac{\partial v^{\beta}}{\partial x^{\alpha}}\right)~,
\eeq
becomes
\beq
\frac{du_{i}}{dt}=\sum^{N}_{j=1}m_{j}(v_{j}^{\alpha}-v_{i}^{\alpha})
\left(\frac{\sigma_{i}^{\alpha
\beta}}{\rho_{i}^{2}}\right)\nabla_{i}^{\beta}W_{ij}~.
\eeq
This symmetric form ensures an exact energy conservation.

The discretized expression of the deformation rate tensor is
\beq
\derpre{\epsilon}^{\alpha
\beta}_{i}=\frac{1}{2}\sum_{j=1}^{N}\frac{m_{j}}{\rho_{j}}
\left((v_{j}^{\alpha}-v_{i}^{\alpha})\nabla_{i}^{\beta}W_{ij}+(v_{j}^{\beta}-
v_{i}^{\beta})\nabla_{i}^{\alpha}W_{ij}\right)~.
\eeq
Thus
\bea
\derpre{\epsilon}_{xx} &=& \sum_{j}\frac{m_{j}}{\rho_{j}}(v_{j}-v_{i})
\frac{\partial W_{ij}}{\partial x_{i}} \nonumber \\
\derpre{\epsilon}_{xy} &=&
\frac{1}{2}\left(\sum_{j}\frac{m_{j}}{\rho_{j}}(v^{x}_{j}-v^{x}_{i})
\frac{\partial W_{ij}}{\partial
x_{i}}+\sum_{j}\frac{m_{j}}{\rho_{j}}(v^{y}_{j}-v^{y}_{i})
\frac{\partial W_{ij}}{\partial y_{i}}\right)~,
\eea
and similarly for $\derpre{\epsilon}_{yy}, \derpre{\epsilon}_{zz},
\derpre{\epsilon}_{xz}$ and $\derpre{\epsilon}_{yz}$.

\subsection{Deformation model}

The model of mechanical deformation that we implement is the simplest
possible, namely a perfect elastic-plastic model, obeying Hooke's law in 
the elastic regime and a suitable plasticity criterion. In the results 
presented in this paper, we will not reach the regime where plasticity 
appears and leave the study of this regime for future work. In the elastic 
regime, the stress tensor reads
\beq
\frac{dS^{\alpha \beta}}{dt}=2\mu(\derpre{\epsilon}^{\alpha \beta}-\frac{1}{3}
\delta^{\alpha \beta}\derpre{\epsilon}^{\gamma \gamma})+S^{\alpha \gamma}
R^{\beta \gamma}+S^{\gamma \beta}R^{\alpha \gamma}
\eeq
where $\mu$ is the shear modulus of the material and $R$ is the tensor of
stress rotations defined by
\beq
R^{\alpha \beta}=\frac{1}{2}\left(\frac{\partial v^{\alpha}}{\partial
x^{\beta}}
-\frac{\partial v^{\beta}}{\partial x^{\alpha}}\right)~.
\eeq
Its discrete SPH approximation is similar to that of the deformation rate
tensor
\beq
R^{\alpha \beta}_{i}=\frac{1}{2}\sum_{j=1}^{N}\frac{m_{j}}{\rho_{j}}
\left((v_{j}^{\alpha}-v_{i}^{\alpha})\nabla_{i}^{\beta}W_{ij}-(v_{j}^{\beta}-
v_{i}^{\beta})\nabla_{i}^{\alpha}W_{ij}\right)~.
\eeq

We need in addition to specify the equation of state, namely the dependence
of the pressure $P = P(\rho, u)$ as a function of the density $\rho$ and the
internal energy $u$. We use the equation of state due to Tillotson [{\it 
Tillotson}, 1962; {\it Benz et al.}, 1994], which works both for expanded 
as well as condensed phases under large pressure or impacts.  This brings 
in the possibility to study the effect of melting or other extreme conditions
that could occur locally under conditions of fast slipping rates. We will
not fully exploit this possibility in the preliminary results presented below. 
We do find a local temperature rise at the level of boundary particles under 
friction but which is insufficient to lead to melting. We thus do not 
incorporate the physics of thermal diffusion and the effect of the internal 
energy is negligible in the regime of our simulations.  The parameters used 
in our simulations are obtained from [{\it Asphaug and Melosh}, 1993]
for typical geological rocks, such as granite, basalt and sandstone. Their
densities are respectively $2.7~g~cm^{-3}$ for granite and basalt and 
$2.3~g~cm^{-3}$ for sandstone. We have not observed significant differences 
in the solid friction for these different materials.

\subsection{Geometry and scaling of the numerical experiments}

We consider the classical friction experiment in which a block of mass $M$
in contact with a solid substrate is submitted to a normal pressure $P$ and 
to a constant horizontal velocity $v$ (see figure 1). We have worked with 
a block of size $0.5 \times 0.375 \times 0.25~cm^3$ while the substrate has 
dimension  $1 \times 0.5 \times 0.25~cm^3$. Three codes are dedicated to 
the construction and positionning of the block and substrate. The first code 
constructs the objects as ensembles of particles. The second code chisels 
the boundaries of the block and substrate, their rugosities and put them 
in contact.  The initial distance and conditions are thus determined. The 
third code calculates the final dimensions of the objects and retrieves the 
coordinates of the particles at the boundaries (first and last layer of 
each object) on which the pressure and velocity conditions are applied.
The particles are put on a regular lattice. We have used the cubic lattice
and the compact hexagonal lattice. In our simulations, we have used between 
$3000$ and $15000$ particles, as a compromise between meshing and computation 
time.  Figure 2 shows a configuration where the particles making up the 
block and substrate are represented\,: in this example, the total number 
of particles in the block is $12 \times 16 \times 20 = 1520$.  Thus the 
size of a particle is of order $0.025$~cm. A slider block of a centimeter 
scale can thus be viewed as being constituted of effective particles of a
fraction of a millimeter. We must thus 
incorporate the correct physics at the scale of each particle.  Each particle at the 
boundary can be viewed as an elementary asperity that will interact
with the particle-asperities of the substrate.

Associated to their size $h$ , mass $m$ and interaction with neighbors,
each particle has a characteristic oscillation frequency $\omega_0$ given by
\beq
\omega_0 = \sqrt{K \over m} \approx {1 \over h}~\sqrt{E \over \rho}~,
\label{rrdsd}
\eeq
where $E$ is the Young modulus and $\rho$ is the density of the material.
We have used the fact that the effective elastic constant $K$ felt by 
a particle is $K \approx h~E$ and the mass of a particle is $m
\approx \rho ~h^3$.
Take $h \approx 0.025~cm$, $E = 10^{11}~Pa$ and $\rho \approx
3~10^3~kg.cm^{-3}$, we get
$\omega_0 \approx 2.3~10^7~s^{-1}$ and a natural frequency $\omega_0/2\pi
\approx 4~10^6~s^{-1}$.
 The expression (\ref{rrdsd}) can be written in terms of the period of
oscillation
 \beq
 T_0 \equiv {2\pi \over \omega_0} = 2\pi~h \sqrt{\rho \over E} = 2\pi~{h
\over c} \approx
 2.5~10^{-5}~s~,
 \eeq
 where $c$ is the longitudinal sound velocity.
This period can be compared to the time
\beq
t_0 = {h \over v}
\eeq
it takes for a particle driven at a velocity $v$ to move over the distance
$h$. $t_0/T_0
= (1/2\pi)~(v/c)$ is thus small for subsonic slider block velocities.

The boundary particles are accelerated due to their collision with the 
substrate under the imposed sliding velocity. Due to their acceleration, 
they entrain their neighbors which themselves accelerate and entrain their 
neighbors, and so on. Macroscopically, this is nothing but wave radiation 
which may act as an important damping process. Since the only physical 
ingredient of our model incorporates elastic interaction, friction
can only emerge as a result of dissipation due to wave radiation. We thus
need to estimate how the efficiency of this radiation is modified by the 
coarse-graining at the scale of the particles and correct for it.

Recall that a particle of mass $m$ submitted to a force
$F_{ext}$ accelerates according to
\beq
m_{i} \derpre{v} = F_{ext}~.
\label{eq-new}
\eeq
Radiation is a generic phenomenon that reflects the acceleration of the
body under consideration. Generically, the power radiated from a body 
accelerating at $\derpre{v}$ is proportional to the square of its acceleration
\beq
P(t) = m T_0 \derpre{v}^{2}~.
\label{eq-tau}
\eeq
Note that this result holds not only for acoustic waves, but for any wave
(electromagnetic, gravitational, hydrodynamics, etc). This radiation exerts 
a feedback force $F_{rad}$ that modifies the acceleration of the body as 
follows [{\it Jackson}, 1962].  Assuming the existence of this radiation 
force, we replace (\ref{eq-new}) by
\beq
m \derpre{v} = F_{ext} + F_{rad}~.
\eeq
$F_{rad}$ is determined from the condition that its work during
$t_{1}<t<t_{2}$ is equal to the radiation energy
\beq
\int_{t_{1}}^{t_{2}} F_{rad} \cdot v dt = - \int_{t_{1}}^{t_{2}} m T_0
\derpre{v}^{2} dt ~.
\eeq
Integrating by part, we get
\beq
\left.
\int_{t_{1}}^{t_{2}} F_{rad} \cdot v dt = m T_0 \int_{t_{1}}^{t_{2}}
\dersec{v}\cdot\derpre{v} dt
-m T_0 (\derpre{v}\cdot v) \right\rbrack _{t_{1}}^{t_{2}}~.
\eeq
For a periodic motion or if $(\derpre{v}\cdot v)=0$ at $t=t_{1}$
and $t=t_{2}$,
\beq
\int_{t_{1}}^{t_{2}} (F_{rad}- m T_0 \dersec{v})\cdot v dt = 0~,
\eeq
thus leading to
\beq
F_{rad} = m T_0 \dersec{v} ~.
\eeq
This effective force is indeed a dissipation as its sign changes under time
reversal $t \to -t$\,: recall that dissipation is nothing but the lack of 
invariance of the motion with respect to the change $t \to -t$.
The important point is that the dissipation force due to radiation is
proportional to the derivative of the acceleration, i.e. to the third-order 
derivative of the position.  This is quite different from the first-order 
derivative dependence of standard fluid friction.  The upshot is that 
radiation is extremely efficient at high frequencies, since its power 
is proportional to $\derpre{v}^{2}$ and thus to the fourth power of the
frequency, according to (\ref{eq-tau}) (this is nothing but Rayleigh's 
scattering law for radiations from objects smaller than the wavelength, 
which is indeed universal as it relies solely on dimensional considerations 
as first derived by Rayleigh (see [{\it Sornette}, 1989] for a review 
and references therein).  The problem however from our perspective is 
that radiation efficiency becomes very small for small frequencies. 
Since coarse-graining using a finite particle size decreases the
natural oscillation frequencies, the resulting radiated power will be
largely reduced compared to the case of a real material in which the particle 
sizes are the atomic scale.

To estimate this effect of coarse-graining on the radiation efficiency, we 
take $\derpre{v} = A \omega_0^2$ with the typical amplitude of motion given 
by $A = h~\epsilon$.  A reasonable estimation of the strain $\epsilon$ 
is such that the elastic potential energy stored per unit volume $(1/2) 
\sigma \epsilon$ be equal to the kinetic energy density $(1/2) \rho v^2$. 
This leads to $\epsilon = v~\sqrt{\rho/E}$ and thus $A = h~v~\sqrt{\rho/E}$. 
Inserting in (\ref{eq-tau}), we find that the radiated power per particle 
is $P = 2\pi~\sqrt{\rho~E}~v^2~h^2$.  Physically, the important quantity 
to establish the energy balance and the friction law is the radiated energy 
per unit volume
\beq
P/h^3 =  2\pi~\sqrt{\rho~E}~{v^2 \over h}~.
\eeq
Since $h$ is about $10^5$ larger than the atomic scale, the radiation
efficiency in our coarse-grained formulation stemming from the acceleration 
of the SPH particles may be down to $10^{-5}$ that of expected physical one. 
This implies that we need to add a dissipation term to the equation in each 
particle to account for the physical radiation due to the accelerations at 
the sub-particle scale that is not correctly described by the coarse-graining 
at the particle scale. Our hypothesis, that will be confirmed below, is that 
friction is not sensitive to the specific form of the internal dissipation 
as long as it is present to damp out the vibrations.

\subsection{Dissipation}

Friction is fundamentally about dissipation. As we already mentionned,
this dissipation can be in the form of radiated waves or conversion of 
local vibrations (phonons) into others modes, thus corresponding to an 
effective loss of energy. To be consistent with our preceding discussion, 
we need to add a dissipation term in the equation of motion proportional 
to the derivative of the acceleration. We would however need to scale the
amplitude of this term by a factor up to $10^5$ in order to obtain a correct 
scaling of the radiation power at the characteristic particle frequency. 
The problem however is that a much larger spectrum of frequencies are 
excited in the complex sliding motion and it is not possible to scale
simultaneously the radiation at all these frequencies simultaneously. We
have prefered a simpler approach which is to add a standard viscous 
dissipation force on each particle
\beq
f_{diss} = -m \gamma v~,
\label{rfgjkl}
\eeq
with a viscous coefficient which is a parameter of the model. Here, $v$ is the
velocity of a particle with respect to the center of mass of the slider block.
As a consequence, this
viscous dissipation acting within each particle damps their motion and thus
converts a part of the sliding kinetic energy into losses that will finally 
produce the solid friction behavior. We have varied $\gamma$ in the interval 
between $0$ and $10^{10}~s^{-1}$. Consistent with the use of this term (\ref{rfgjkl})
as a device to mimick losses from radiation processes, we have not included this term
in the energy balance equation. In other words, any heat produced by this dissipation is assumed
to be instantaneously radiated.

Before resorting to this viscous dissipation, we have explored various other
possibilities, such as trapping the acoustic waves in dissipative cavities
of various form, so as to mimick out-going escaping radiated waves that 
never come back.  Unfortunately, in addition to the partial backscattering 
occuring at junctions (even with our best effort to adapt the acoustic 
impedance by using smooth geometries and slowly varying mechanical properties), 
we found that the pattern of the acoustic particle vibrations self-organized 
so that the amplitude became vanishing small at the borders of these traps, 
making them totally inefficient. Recall that during a typical experiment of 
one millisecond at $10~m/s$, the slider block slips over $1~cm$ while
the acoustic vibrations propagate over about $5~m$, i.e. have time to make
$5000$ travels back and forth within the slider block. There is a lot of 
shaking and organization of particle vibration going on all the time!

From a numerical point of view, it is also necessary to incorporate a numerical 
viscosity that allows to regularize large gradients as occur for instance 
in shocks. This artificial viscosity modifies the equation of momentum 
conservation into
\beq
\frac{dv_{i}^{\alpha}}{dt}=\sum^{N}_{j=1}m_{j}\left(\frac{\sigma_{i}^{\alpha \beta}}{\rho_{i}^{2}}
+\frac{\sigma_{j}^{\alpha
\beta}}{\rho_{j}^{2}}+\Pi_{ij}\right)\nabla_{i}^{\beta}W_{ij}~.
\label{mom-vis}
\eeq
Correspondingly, the artificial viscosity also brings in a contribution to
the equation of energy conservation. The addition term is similar to a pressure
\bea
\Pi_{ij}=\left\{
\begin{array}{cc}
\frac{-\alpha \bar{c}_{ij}\mu_{ij}+\beta \mu^{2}_{ij}}{\bar{\rho}_{ij}}
&\mbox{if\ }(\vec{v}_{i}-\vec{v}_{j})\cdot(\vec{r}_{i}-\vec{r}_{j})\leq 0 \\
0 & \mbox{else~,}
\end{array}
\right.
\eea
where $\bar{c}_{ij}=\frac{1}{2}(c_{i}+c_{j})$ and $c_{i}$ and
$c_{j}$ are the sound velocities in the particles $i$ and $j$. The average
density is defined by $\bar{\rho}_{ij}=\frac{1}{2}(\rho_{i}+\rho_{j})$. The
estimation of the divergence of the relative velocity between $i$ et $j$ is
\beq
\mu_{ij}=\frac{h(\vec{v}_{i}-\vec{v}_{j})\cdot(\vec{r}_{i}-\vec{r}_{j})}
{|\vec{r}_{i}-\vec{r}_{j}|^{2}+\epsilon h^{2}}~.
\eeq
The addition term $\epsilon h^{2}$ prevents a divergence occurring for small
 $|\vec{r}_{i}-\vec{r}_{j}|$. For instance, the choice $\epsilon=0.01$ gives a
 correct smoothing of the velocity only if the interparticle distance is
larger than
$0.1h$.

Note that $\Pi_{ij}$ remains symmetric in $i$ and $j$ which ensures the
conservation of the total linear and angular momenta. The artificial viscosity 
disappears when the two particles separate, which ensures that this dissipation 
obeys the second law of thermodynamics, i.e. can only increase the entropy 
of the system.  The values of the parameters $\alpha$ and $\beta$ are not 
critical. In the following, we adopt $\alpha=1$ and $\beta=0.1$. We have 
checked that the additional ``dissipation'' stemming from rounding and 
integrating errors is negligible. 

This artificial viscosity introduced in the SPH formulation
may produce a non-negligible dissipation when shear is important
and we have been concerned with the possibility that the effects reported below
could stem purely from numerical effects. We believe that the artificial 
viscosity does not produce in itself a solid friction from three observations\,:
\begin{itemize}
\item the relative velocities of the particles with respect to the center of mass
are small\,;
\item the solid friction coefficient is zero, to within numerical accuracy, when
$\gamma = 0$, i.e. the artificial viscosity alone does not produce a detectable
friction\,; 
\item the temperature does not increase appreciably. This provides an upperbound
for the dissipated energy due to the artificial viscosity (recall that the 
viscosity (\ref{rfgjkl}) is not incorporated in the heat production as mentionned above),
which is thus found negligible.
\end{itemize}
We conclude that the results reported below are
not the expression of numerical tuning.

\subsection{Implementation}

Each particle carries $13$ variables\,: three positions $(x, y, z)$, three
velocities $(v_x, v_y, v_z)$, the density, the energy, and five components 
of the deviatoric stress tensor $s_{xx}, s_{yy}, s_{xy}, s_{xz}, s_{yz}$. At each 
time step, this set of $13$ variables is incremented. The temporal integration 
is carried out using a second order Runge-Kunta-Fehlberg algorithm with 
adaptative time steps. To ensure second order accuracy, the forces are 
evaluated twice per time step. When a particle is suddenly colliding with 
another one, the local stress can jump to high values resulting in a 
drastic reduction of the time step reaching in some cases a factor $100$!
The typical time step is $5~10^{-8}$ seconds. Our simulations are run over
$20,000$ to $50,000$ time steps, i.e. over a total duration of a millisecond 
or more.  The simulations have been performed on RISK6000 and Ultra-Sparc 2 
workstations and typically last three days.

Starting from a configuration where the block is in contact with the substrate, 
at a distance equal to the inter-particle spacing, we submit the block to a
vertical normal pressure along the $z$ axis. This is done by applying the 
same force on all the particles on the upper boundary of the block. The 
bottom boundary of the substrate is fixed and cannot move.  The pressure 
is applied progressively as $1-\exp(-t/\tau)$, with a characteristic
$\tau \approx 30-250~\mu s$, which is not too short to allow for the 
propagation of acoustic waves back and for the system to reach mechanical 
equilibrium. We have varied the pressure from $10^6~Pa$ to $10^9~Pa$. 
Once the equilibrium is {|bf reached, we impose a horizontal sliding
velocity $v$. This velocity was varied $v$ in the range $0.1$ to $10~m/s$.
During the first integration time step, all particle in the sliding 
block are moved with velocity $v$. During the rest of the integration, only
those particles in the first and sometimes second layers of the sliding 
block are moved with velocity $v$ the others ones evolving according to
the laws of elasticity. This somewhat cumbersome starting procedure 
prevents large unwanted inertial oscillations to occur that would require 
a long time to damp and brings the slider block efficiently to a constant 
velocity.} Our choice to push the sliding block at its upper boundary is 
an attempt to mimick large scale driving boundary conditions. It also
allows the block to freely adjust itself close to the interface with the
 substrate in the hope to minimize the influence of boundary conditions.

At each time step, the forces exerted on each particle are calculated. They
are the cohesion force $f_{coh}$ between each particle, the uniaxial 
pressure $f_{pb}$ exerted at the upper boundary of the slider block, 
the repulsion force $f_{spr}$ at the block-substrate interface that acts only
between the particles in the first layer of each objects, the viscous
dissipation force $f_{dis} = -\gamma (v_{x}-v)$ along the $x$-axis, 
and $-\gamma v_{y}$ along the $y$-axis and $-\gamma v_{z}$ along the $z$-axis. 
The conservation of momentum reads
\beq
m \derpre{v} = f_{coh} + f_{pb} + f_{spr} + f_{dis}~.
\eeq
The solid friction force is measured as the force  $f^x_{spr}$ exerted
along the direction of motion $Ox$ by the substrate on the particles
in the first layer of the slider block in contact with the substrate. We
also measure the vertical component and the ratio
\beq
\mu \equiv f^x_{spr}/f^z_{spr}
\label{rfchhjkx}
\eeq
gives a local and instantaneous measure of the solid friction coefficient. 
We can then perform a time average and space average to get the macroscopic 
friction coefficient.

We have investigated several types of boundary forces between substrate and
slider block particles\,:
\begin{enumerate}
\item spring-like force
\bea
f_{1}(r) = \left\{
\begin{array}{cc}
-K(r-h) & 0<r<h \\
 0 & r>h
\end{array}
\right.
\label{eq-f1}
\eea

\item Smooth repulsive force [{\it Monaghan}, 1994]
\bea
f_{2}(r)=\left\{
\begin{array}{cc}
A (\frac{l}{r})^{s} & 0<r<l \\
D (r-b)^{2} & l<r<b \\
0           & r>b
\end{array}
\right.
\label{eq-f2}
\eea
This force is also radial and its first derivative is continuous. The
continuity condition at $r=l$ gives $D = A (\frac{s+2}{2b})^{2}$ and 
$l = \frac{sb}{(s+2)}$. In our calculations, we take $b=h$.

\item Lennard-Jones force per unit of mass
\bea
f_{3}(r) = \left\{
\begin{array}{ccc}
\epsilon \left((\frac{l}{r})^{m}-(\frac{l}{r})^{n}\right) & 0<r<r_{c} &
\mbox{with\ } m>n \\
A (R-r)^{2} + D (R-r) & r_{c}<r<R \\
0  & r>R
\end{array}
\right.
\label{eq-f3}
\eea
$f_{3}(r)$ vanishes for $r=l$ and $r=R$ and has a minimum at $r=r_{c}$.
In our simulations, we have taken $R=\frac{3}{2}\Delta p$ and $l=\Delta p$.
The continuity of $f_{3}(r)$ and of its first derivative at
$r=r_{c}$ implies
\beq
l = r_{c}\left(\frac{n}{m}\right)^{\frac{1}{m-n}}
\eeq
and
\beq
A = - \frac{D}{2(R-r_{c})}~,
\eeq
from which we derive
\beq
D = \frac{2\epsilon}{R-r_{c}} \left[
\left(\frac{n}{m}\right)^{\frac{m}{m-n}} -
\left(\frac{n}{m}\right)^{\frac{n}{m-n}} \right] ~.
\eeq
$\epsilon$ remains an adjustable parameter that is chosen so that the time
scale
\beq
\delta t = \frac{1}{2}\frac{h}{c_{s}}~.
\eeq
With $\Delta p =h$ , this leads to
\beq
\epsilon =
\frac{4}{9}\frac{c_{s}^{2}}{h}\left(\frac{n}{m}\right)^{\frac{1}{m-n}}\frac{1}{m
-n}
\eeq
We have used $m=8, n=4$  and $m=12, n=6$ and found no significant
difference in the results.

\end{enumerate}

Even without introducing asperities explicit, the potential field seen 
by particles at the interface between substrate and slider block is not
smooth. Due to the particle structure, the potential thus exhibits a periodic 
structure of troughs and peaks with a period equal to that of the cubic or 
hexagonal lattice used to construct the block and substrate. This corresponds to
a slider block with a periodically modulated roughness sliding on a substrate
presenting also a periodically modulated roughness. We have also investigated
cases for which we explicitly introduced random asperities. 

\section{Solid friction for periodic roughness}

\subsection{Measurements}

We have carried out simulations for various dissipation coefficient
$\gamma$. Figure 3a shows the local friction coefficient $\mu \equiv 
f^x_{spr}/f^z_{spr}$ as defined in (\ref{rfchhjkx}) measured on a 
single particle in the first layer of the slider block in contact with the 
substrate. The simulation uses the boundary force $f_{1}$ defined by 
(\ref{eq-f1}). The total number of particles used is $5504$, the sliding 
velocity $v = 1~m/s$, the applied pressure $10^{8}$ Pa, and the dissipation
parameters $\alpha=0.1, \gamma=10^{6}~s^{-1}$. The positive and negative 
oscillations correspond to alternative braking and accelerating phases of 
the particle as it climbs up and down the asperities of the substrate. 
The time average of this local instantaneous friction coefficient is $\mu 
\approx -0.05$, the negative sign corresponding to a net drag. Figure 3b 
shows the global instantaneous friction $\mu_l$, obtained by taking the 
ratio of the total force along $x$ on all block particles on the boundary 
in contact with the substrate to the total force along $z$. We obtain the
same estimate $\mu_l \approx -0.05$ for the friction coefficient after 
time averaging.

The following table summarizes our results for the effective friction
coefficient for three different values of the dissipation $\gamma$. The 
simulations have otherwise been carried out exactly in the same way, 
using $5504$ particles, an imposed sliding velocity of $1~m/s$ and an 
applied pressure of $10^8~Pa$. In the Table, $\delta \mu$ denotes
the standard deviation of $\mu$, i.e. the amplitude of its fluctuations
\bc
\begin{tabular}{cccc}
$\alpha$ & 0 & 0,1 & 0,1 \\
$\gamma$ & 0 & 0 & $10^{6}$ \\
$\mu$ & -0.020 & -0.016 & -0.058 \\
$\delta \mu$ & 0.017 & 0.017 & 0.007\\
$\mu_{l}$ & -0.021 & -0.031 & -0.050 \\
\end{tabular}
\ec
These results are compatible with a vanishing average friction in absence
of internal dissipation. When internal friction is present, the erratic 
motions of the particles submitted to damped multiple acoustic paths decrease 
somewhat and produce a finite friction coefficient which is significantly 
smaller than the instantaneous fluctuations. Interestingly, the value of
of the friction coefficient does not appear to change with increasing $\gamma$.
Above $\gamma = 10^9~s^{-1}$, the numerical time step becomes so small 
making the calculation almost impossible and thus preventing us from 
studying this regime. For $\gamma < 10^6~s^{-1}$, the fluctuations are 
too great to allow for accurate measurements. To summarize, we obtain
\beq
\mu \approx \mu_l \approx -0.06 \pm 0.01~, ~~~~~~~~~{\rm for}~~
10^6~s^{-1} \leq \gamma \leq  10^9~s^{-1}~.
\eeq
This result justifies a posteriori our procedure to model the radiation
damping by the viscous dissipation (\ref{rfgjkl}). The remarkable fact that 
the solid friction coefficient is independent by and large of the amplitude 
of damping suggests the following picture: in the elastic regime we explore,
solid friction stems from the acceleration of asperities brought in contact 
and collision that radiates high-frequency vibrational waves subsequently
damped out, thus converting a part of the kinetic energy of the slider block 
into dissipation. The specific form of the internal dissipation seems not 
to be important, as long as there is a dissipation that can damp out the 
vibrations of the asperities.

\subsection{Particle motions}

Figure 4 shows the vertical motion of a particle of the slider block in the
layer in contact with the substrate, using the boundary force $f_{1}$, $5504$
particles, $v = 1~m/s$, a pressure of $10^{8}~Pa$, $\alpha=0.1, 
\gamma=10^{6}~s^{-1}$. The climb and fall over the particle-asperities of 
the substrate are clearly visible. The downward drift is caused by a steady 
horizontal drift of the slider block along the $y$ direction, perpendicular 
to the driving velocity. The horizontal motion of such a typical particle is 
almost steadily increasing, with however a short stop just when the particle 
is at the bottom of the potential created by the substrate asperities.

Figure 5 shows the velocity along $x$ (fig. 5a) and along $z$ (fig.5b)
of the center of mass of the slider block, under the same conditions as 
described in figure 4. This shows that the slider block as a whole moves 
up and down as well as periodically accelerates and decelerates due to 
the interactions with the periodic array of asperities of the substrate.

For small $\gamma \sim 10^6~s^{-1}$, the slider block exhibits rather large
vertical oscillations that decrease significantly in amplitude as $\gamma$ 
increases. For the largest explored $\gamma$, the slider block follows very 
closely the geometry of the substrate asperities. Note that the motion of 
a particle in the layer in contact with the substrate is a very good proxy 
for the motion of the slider block as a whole.

\subsection{Pressure and velocity dependence}

For pressure below $10^7~Pa$, we are unable to measure a non-zero friction
coefficient. Independently of the dissipation level $\gamma$, the measured 
$\mu$ are within uncertainty the same with or without dissipation. The 
reason is that the asperities do not much penetrate into each other and 
the slider block ``floats'' over the substrate without generating significant 
vibrations that can be dissipated. While more simulations are required to
demonstrate it, we surmise that this behavior is due to the existence in
the elastic regime of a critical pressure threshold below which there is
zero friction. 

For pressure above $10^9~Pa$, the substrate force $f_{1}$ cannot be used
anymore as the slider block penetrates within the substrate. We have then 
used $f_{2}$ and $f_{3}$ given by (\ref{eq-f1}) and (\ref{eq-f1}) respectively. 
Unfortunately, the integration time step shrinks drastically, thus limiting 
an exhaustive exploration of this regime.  However, we have found that 
the results are the same for a pressure of $10^9~Pa$ as found for $10^8~Pa$. 
We have also verified that the three different forces give the same results.

With respect to the velocity dependence, our time-explicit numerical SPH
method does not allow us to explore too {\it small} velocities due to
the prohibitive calculation time. A few runs at $v = 0.1~m/s$ give essentially 
the same results as for $v = 1~m/s$. However, for larger velocities $v = 
10~m/s$, the friction coefficient increases and doubles at $\mu = 0.11$. 
The instantaneous value fluctuates with much larger amplitudes as a result 
of a very high level of vibrational excitations induced by the collision 
between asperities.  The corresponding instantaneous friction coefficient
measured on all the particles in the boundary layer in contact with the
substrate is shown in figure 6. The simulation uses the boundary force 
$f_{1}$ and has $5504$ particles.  The parameters are $v = 10 ~m/s$, which 
was imposed at $t = 0.2~ms$, a pressure of $10^{8}~Pa$ and dissipation 
$\alpha=0,1, \gamma=10^{6}~s^{-1}$. Note the existence of flat steps 
in the graph at the value $\mu = 0$\,: they correspond to jumps of the 
slider block over the substrate asperities. In these regimes, the slider 
block is literally flying over the substrate, as a result of an efficient
transformation of horizontal to vertical momentum induced by the collisions 
with the substrate asperities. This regime has been postulated first by 
Lomnitz-Adler [1991] and our simulations confirm nicely his ideas.
As a result of these jumps, the landing of the slider block does
not occur in phase with the substrate. As a consequence, the evolution 
becomes chaotic, in the rigorous mathematical meaning of the term. The mechanism 
for this chaotic behavio  is similar to that in the toy model of a bouncing 
ball on a sinusoidally vibrating table [{\it Mehta and Luck}, 1990; 1993; 
{\it Franaszek and Isomaki}, 1991; {\it Luo and Han}, 1996; {\it de Oliveira 
and Goncalves}, 1997].  The jump of the slider block occurs from roughly 
the maximum of the potential created by the substrate asperities and over 
its descent, i.e. in the pulling portion of the potential. This explains 
why the total time average friction coefficient is stronger as the pulling 
part as become weaker. We have not push more the numerical exploration of 
this very interesting behavior and leave it to a future publication.
We expect even more interesting behavior at still larger velocities as the
jump can carry the slider block over two or more asperities leading to the
possibility of a rich phenomenology for the friction coefficient at these 
high sliding velocities.

To summarize, the main result of our investigation of the velocity
dependence of the solid friction coefficient is that it {\it increases} at 
large velocities. This is in agreement with the expectation that the vibrational 
radiation damping becomes the dominating mechanism with an efficiency that 
increases fast with the velocity.

\section{Disordered and fractal interfaces}

We have investigated three types of disordered roughness\,:
incommensurate periodic roughness between the block and substrate, a step
and random roughnesses.

\subsection{Commensurate periodic roughness}

Nothing changes compared to the previous periodic case, except for the fact
that we tilt the lattice structure of the slider block with respect to the 
substrate by an angle between $0$ (previous case) and $45$ degrees. When 
the angle is non zero, the asperities of the slider block do not encounter 
those of the substrate in the same configuration and at the same time.  
We find that, for most angles, the friction coefficient is slightly
decreased compared to the value $0.06$ of the periodic case. We find a
remarkable result only for the special case of the most incommensurate regime 
where the ratio of the mesh size of the slider block to that of the substrate 
lattice is equal to the golden mean $(\sqrt{5} + 1)/2 \approx 1.618$. Recall 
that the golden mean is the irrational number that is the least well 
approximated by a rational number. For this ratio, no two asperities of the 
slider block will be in the same configuration at the same time with respect 
to an asperity of the substrate. The measured friction coefficient at $v =
1~m/s$, a normal pressure of $10^8~Pa$ and $\gamma = 10^6~s^{-1}$ is extremely
small \,: $\mu_l = -0.003$. Qualitatively, we attribute this small value to 
the conjunction of two effects. First, 
the tuning of the vibrational resonances occurring in the perfect periodic 
case does not occur anymore. This leads to much smaller coherent vibrations 
and thus smaller damping. Secondly, the still regular smooth roughness does not produce 
large local amplitudes of vibrations. We are nevertheless surprised that these 
effects contribute to such a small value of $\mu_l$.  Our result is reminiscent 
of the ``super-lubrification'' regime found recently [{\it Hirano et al.}, 
1997] using a tungsten $W_{(011)}$ wire sliding over a silicium $Si_{(001)}$ 
surface in an incommensurate geometry and also over solid $MoS_2$ solid films 
[{\it Martin et al.}, 1993].

\subsection{Step-like roughness}

The substrate is made of a ``smooth'' periodic surface up to some fixed
$x_{step}$, at which a vertical step equal to the particle size $h$ is made 
by adding a single particle layer beyond $x_{step}$.  The simulations are 
performed as before. The slider block is accelerated at $v$ before the
step. The simulations use the substrate force $f_{1}$, $5328$ particles
forming cubic lattices.  Figure 7 corresponds to $v = 1~m/s$ and a normal 
pressure of $10^6~Pa$. The arrows represent the instantaneous velocities of 
the particle. Figure 7a shows a snapshot exactly at the time when the slider 
block encounters the step. Figure 7b shows that the slider block is ejected 
vertically and starts to jump over the step. Figure 8 corresponds to $v = 
10~m/s$ and a normal pressure of $10^8~Pa$. It shows a latter stage when 
the slider block is in flight above the substrate. The arrows now show the 
stress carried by each particle projected in the 2D $(x,z)$ plane. As the 
slider block flies over the substrate, the stress within it relaxes to zero. 
Its landing occurs several particles after the step and the slider block is 
found to bounce back several time before resumming its steady state sliding.
These simulations demonstrate again the importance of jumps at high velocities, 
even in the presence of strong confining pressure.

\subsection{Random roughnesses}

Three types of randomness have been investigated: holes in the first layer 
of the subtrate, variable heights of the substrate particles in the first 
layer and fractal roughness.

\subsubsection{Holes in the first layer of the substrate}

One could imagine first to introduce disorder by removing at random a
fraction of the particles in the first layer of the substrate in contact with 
the slider block, thus creating holes of varying sizes and shapes controlled 
by the distribution of cluster sizes in 2D percolation [{\it Stauffer and 
Aharony}, 1994]. It turns out that, for a density of holes no larger than
$60~\%$, this has no effect on the slider block as it continues to be
supported by the remaining particles of the substrate. We find the same 
coefficient of friction as in absence of holes. Above $60~\%$, the slider 
block starts falling partly in sufficiently large holes and the regime of 
sliding becomes controlled by the jumps over steps as just described.
This value of $60~\%$ corresponds approximately to the concentration of
holes at which the large clusters in the substrate become of size comparable 
to the slider block.

\subsubsection{Variable heights of the substrate particles in the first layer}

The substrate is now made of a single layer of particles. These particles
are again positionned regularly on a lattice in the $x-y$ plane, but their 
vertical positions are taken randomly and uniformly between $-\Delta z/2$ 
and $+\Delta z/2$. The slider block is not modified. The solid friction 
coefficient is now measured by measuring the total force exerted on the 
first layer of the slider block in contact with the substrate, as the 
particles in the first layer of the slider block are not continuously in 
contact with the substrate due to its random roughness.  Figure 9 shows 
the measured friction coefficient $\mu_l$ as a function of time for a
simulation using $f_{1}$, $6640$ particles forming a cubic lattice, 
$v = 1~m/s$ and a normal pressure of $10^8~Pa$.  After an initial large 
resistance at the beginning of the slider block motion, the friction
coefficient settles to a stationary regime characterized by random
fluctuations still decorated by a periodic structure reflecting that of 
the asperities of the slider block.  We find that $\mu_l$ increases with 
the roughness amplitude $\Delta z$ of the substrate.  For $\Delta z = 0.5~h$, 
$\mu_l = - 0.017$ and increases continuously to  $\mu_l = - 0.08$ for 
$\Delta z = h$. Surprisingly, a small roughness {\it decreases} the solid 
friction while a larger roughness increases it above its periodic
roughness value $0.06$. We attribute the decrease of $\mu_l$ for small 
$\Delta z$ to the detuning of the vibrational resonances occurring in 
the perfect periodic case, that were at the origin of relatively large 
vibrational radiation and thus damping.

\subsubsection{Fractal roughness}

As in the previous section, the substrate is made of a single layer of
particles, positionned regularly on a 2D cubic lattice in the $x-y$ plane 
with mesh $h$.  Their vertical positions are determined by using the spectral 
synthesis method described in [{\it Peitgen and Saupe}, 1988] to generate 
a self-affine surface. We have investigated different dimensions between 
$D_f = 2.1$ to $D_f = 2.9$. The slider block is made of $6440$ particles 
organized in a hexagonal compact lattice. Thus, even without the fractal
vertical structure of the substrate, there is no more commensurability
between the slider block and substrate. The maximum amplitude of the 
self-affine surface is imposed equal to $h$ for all values of $D_f$. The 
largest wavelength that we have kept in the construction of the substrate 
is equal to one eighth the length of the slider block. This ensures that 
the slider block remains stable and does not fall or jump over steps as in 
the step case.  Figures 10a and 10b show two fractal surfaces, respectively 
with $D_f = 2.3$ and $D_f = 2.8$, with the same maximum amplitude $h$.  
Figures 11a and 11b show the locii of contacts between the substrate and 
the slider block for the two surfaces shown in figure 10. Figure 11a, 
corresponding to $D_f = 2.3$, shows a larger and more coherent contact area 
than figure 11b, corresponding to $D_f = 2.8$.

Figure 12 shows the variation of the solid friction coefficient $\mu_l$ as
a function of $D_f$ for simulations performed under a normal pressure of 
$10^8~Pa$, $v= 1~m/s$ and $\gamma = 10^6~s^{-1}$. For $D_f < 2.6$, the 
friction coefficient is found less than that of the perfect periodic case. 
As for the previous case, we attribute this result to the detuning of the
vibrational resonances occurring in the perfect periodic case, that were at
the origin of relatively large vibrational radiation and thus damping.
For $D_f > 2.6$, the friction coefficient becomes larger than that of
the perfect periodic case, reflecting the transition to another regime
controlled by the larger elastic distorsion of the asperities.

\section{Conclusion}

We have presented a new numerical model for the investigation of solid 
friction properties in the regime of fast relative velocities of the order 
of meters to tens of meters per second. We have restricted our investigation 
to the regime where only elastic deformations occur between the asperities 
at the contact between the slider block and the substrate. In this case, 
the only mechanism that dissipates energy and creates a non-vanishing solid 
friction coefficient is through the generation of vibrational radiations 
that are subsequently damped out, either by escaping to infinity
or by a suitable internal damping process. We have examined periodic
commensurate and incommensurate asperities and various types of disordered 
surface. In this elastic regime, we report the evidence of a transition 
from zero (or non-measurable) friction to a finite friction when the 
normal pressure increases above about $10^6~Pa$.  We find a remarkably
universal value for the friction coefficient $\mu \approx 0.06$, which is
independent of the internal dissipation strength over three order of 
magnitudes, and independent of the detailled nature of the slider 
block-substrate interaction. We find that disorder may either decrease 
or increase $\mu$ due to the competition between two effects: Disorder 
detunes the coherent vibrations of the asperties that occur in the 
periodic case, leading to weaker acoustic radiation and thus weaker 
damping. On the other hand, large disorder leads to stronger vibration 
amplitudes at local asperities and thus stronger damping.  Our simulations 
have confirmed the existence of jumps of the slider blocks that occur over 
steps or asperities for the largest velocities of $10~m/s$. We find a
velocity strengthening with a doubling of the friction coefficient when 
the velocity increases from $1~m/s$ to $10~m/s$. This reflects the 
increasing amplitude of vibrational damping.

We leave for another a later paper the investigation of the regime where 
the pressure is larger and the roughness is more disordered so that the 
local pressure at asperities reaches the plasticity threshold.  In this 
regime, temperature diffusion must be added to the formulation.  This does 
not pose any conceptual difficulty and can easily be incorporated in our code.
In this regime, both vibration damping and plasticity become the source
of dissipation. It is probable that the friction coefficient will be found
larger in this case, as often measured in macroscopic experiments that 
work in the regime where asperities are deformed in the plastic regime 
[{\it Dieterich and Kilgore}, 1994].  However, we still expect that 
jumps play an important role at the largest velocities of tens of meter 
per second.

\vskip 0.5cm
We acknowledge stimulating discussions with T. Villemin and J.-P. Vila.

\vskip 1cm
{\bf References}
\vskip 0.5cm

Asphaug, E. and H.J. Melosh, The stickney impact of phobos: A dynamical model,
Icarus, 101, 144-164, 1993.

Asphaug, E., S.J. Ostro, R.S. Hudson, D.J. Scheeres and W. Benz,
Disruption of kilometre-sized asteroids by energetic collisions,
Nature, 393, N6684, 437-440, 1998.

Baumberger, T., and L. Gauthier, Creeplike relaxation at the interface
between rough solids under shear, J. Phys. I France, 6, 1021-1030, 1996.

Beeler, N.M., T.E. Tullis, M.L. Blanpied and J.D. Weeks,
Frictional behavior of large displacement experimental faults,
J. Geophys. Res., 101, 8697-8715, 1996.

Beeler, N.M., T.E. Tullis and J.D. Weeks,
The roles of time and displacement in the evolution effect in rock
friction, Geophys. Res. Lett., 21, 1987-1990, 1994.

Beeman, M., W.B. Durham and S.H. Kirby, Friction of ice,
J. Geophys. Res., 93, 7625-7633, 1988.

Benz, W., Smooth particle hydrodynamics: A review,
The Numerical Modeling of Nonlinear Stellar Pulsations,
J. Buchler, ed., 269-288, Kluwer Academic Publishers, Dordrecht, 1990.

Benz, W.,  W.L. Slattery and A.G.W. Cameron,
The origin of the moon and the single-impact hypothesis,
Icarus, 66, 515-535, 1986.

Benz, W., and E. Asphaug, Impact simulations with fracture. I- Method and
tests,
Icarus, 107, 98-116, 1994.

Benz, W.,  E. Ryan and E. Asphaug, Numerical simulations of catastrophic
disruption: Recent results, In The 4th International Workshop on the
Catastrophic Disruption
of Small Solar System Bodies, Gubbio, Italy, 1994.

Benz, W., and E. Asphaug, Simulations of brittle solids using smooth particle
hydrodynamics, Computer Physics Communications, 87, 253-265, 1995.

Benzion, Y., and J. Rice, Earthquake failure sequences along a cellular
fault zone in
a 3-dimensional elastic solid containing asperity and nonasperity regions,
J. Geophys. Res., 93, 14109-14131, 1993.

Benzion, Y., and J. Rice, Slip patterns and earthquake populations
along different classes of faults in elastic solids, J. Geophys. Res., 100,
12959-12983, 1995.

Bocquet, L., and H.J. Jensen, Phenomenological study of hysteresis
in quasistatic friction, J. Physique I, 7, 1603-1625, 1997.

Bowden, F.P., and D. Tabor, The Friction and Lubrication of Solids
(Oxford University Press, 1954).

Brace, W.F., Laboratory studies of stick-slip and their application to
earthquakes, Tectonophysics, 14, 189-200, 1972.

Brace, W. R., and J.D. Byerlee,  Stick-slip as a mechanism for earthquakes,
Science, 153, N3739, 990-992, 1966.

Brune, J.N., S. Brown and P.A. Johnson,
Rupture mechanism and interface separation in foam rubber models of
earthquakes -  A
possible solution to the heat flow paradox and the paradox of large
overthrusts,
Tectonophysics, 218, 59-67, 1993.

Bullard, E.C., The interior of the earth, pp.57-137, in The Earth as a
planet'', G.P. Kuiper, ed., University of Chicago Press, 1954 (see pp.
120-121).

Byerlee, J. D., et al., Stick slip, stable sliding, and
earthquakes - Effect of rock type, pressure, strain rate, and stiffness,
J. Geophys. Res., 73, 6031-6037, 1968.

Caroli, C., and P. Nozi\`eres,  Dry friction as a hysteretic elastic response,
In Physics of Sliding Friction, Persson et Tosatti, eds., Kluwer
Academic Publishers, 1996.

Caroli, C., and B. Velicky,
Dry friction as an elasto-plastic response: Effect of compressive
plasticity, J. Physique I, 7, 1391-1416, 1997.

Cochard, A., and Madariaga R., Dynamic faulting under rate-dependent
friction, Pure and Applied Geophysic, 142, 419-445, 1994.

Coleman, C.S. and G.V. Bicknell, Jets with entrained clouds - Hydrodynamics
simulations and
magnetic field structure, MNRAS, 214, 337-355, 1985.

Cox, S.J.D., Velocity-dependent friction in a large direct shear experiment on
gabbro, In Deformation Mechanisms, Rheology and Tectonics, Knipe et Rutter,
eds.,
vol. 54, 63-70, Geological Society Special Publication, 1990.

de Oliveira, C.R., and P.S. Goncalves,
Bifurcations and chaos for the quasiperiodic bouncing ball, Phys. Rev. E,
56, 4868-4871, 1997.

Dieterich, J.H., Time-dependent friction in rocks,
J. Geophys. Res., 77, 3690-3697, 1972.

Dieterich, J.H., Time-dependent friction and the mechanics of stick-slip,
Pure and Applied Geophysics, 116, 790-806, 1978.

Dieterich, J.H., Modeling of rock friction: 1. experimental results and
constitutive equations, J. Geophys. Res., 84, 2161-2168, 1979.

Dieterich, J.H., Earthquake nucleation on faults with rate and state-dependent
strength, Tectonophysics, 211, 115-134, 1992.

Dieterich, J., and Kilgore B.D., Direct observation of frictional
constacts- New insight for state-dependent properties, Pure and Applied
Geophysics,
143, 283-302, 1994.

Franaszek, M., and H.M. Isomaki, Anomalous chaotic transients and repellers of
the bouncing-ball dynamics, Phys. Rev. E, 43, 4231-4236, 1991.

Gu, Y., and T.-F. Wong, Effects of loading velocity, stiffness, and inertia
on the
dynamics of a single degree of freedom spring-slider system,
J. Geophys. Res., 96, 21677-21691, 1991.

Henyey, T.L., and G.J. Wasserburg, Heat flow near major strike-slip
faults in California, J. Geophys. Res., 76, 7924-7946, 1971.

Herrmann, H.J., G. Mantica and D. Bessis, Space-filling bearings,
Phys.Rev.Lett., 65, 3223-3226, 1990.

Hirano, M., K. Shinjo, R. Kaneko and Y. Murata,
Observation of superlubricity by scanning tunneling microscopy,
Phys. Rev. Lett., 78, 1448-1451, 1997.

Jackson, J.D.,  Classical electrodynamics, 2d ed., New York, Wiley, 1975.

Jensen, H.J., Y. Brechet and B. Doucot,
Instabilities of an elastic chain in a random potential,
J. Phys. I France, 3, 611-623, 1993.

Johansen, A., P. Dinon, C. Ellegaard, J.S. Larsen et H.H. Rugh,
Dynamic phases in a spring-block system , Phys. Rev. E, 48, 4779-4790,1993.

Knopoff, L., The organization of seismicity on fault networks,
Proc. Nat. Acad. Sci. USA, 93, 3830-3837, 1996.

Lachenbruch, A.H., and Sass J.H., Heat flow and energetics of the San
Andreas fault zone, J. Geophys. Res., 85, 6185-6222, 1980.

Lattanzio, J.C., J.J. Monaghan, H. Pongracic and M.P. Schwarz,
Interstellar cloud collisions, MNRAS, 215, 125-147, 1985.

Lomnitz-Adler, J., Model for steady state friction,
J. Geophys. Res., 96, 6121-6131, 1991.

Luck, J.-M., and A. Mehta, Bouncing ball with a finite restitution -
Chattering,
locking and chaos, Phys. Rev. E, 48, 3988-3997, 1993.

Lucy, L.B., A numerical approach to the testing of the fission hypothesis,
Astron. J., 82, 1013-1024, 1977.

Luo, A.C.J., and R.P.S. Han, The dynamics of a bouncing ball with a
sinusoidally
vibrating table, Nonlinear Dynamics, 10, 1-18, 1996.

Martin, J.M.,  C. Donnet, T. Le Mogne and T. Epicier,
Superlubricity of molybdenum disulphide, Phys. Rev. B, 48, 10583-10586, 1993.

Melosh, H.J., Dynamical weakening of faults by acoustic fluidization,
Nature, 379, 601-606, 1996.

Mehta, A., and J.-M. Luck, Novel temporal behavior of a nonlinear dynamical
model
system - The completely inelastic bouncing ball, Phys. Rev. Lett., 65,
393-396, 1990.

Monaghan, J.J.,  and R.A. Gingold,
Shock simulation by the particle method, J. Comput. Phys., 52, 374-389, 1983.

Monaghan, J.J., An introduction to SPH,
Computer Physics Communications, 48, 89-96, 1988.

Monaghan, J.J., Smooth particle hydrodynamics,
Ann. Rev. Astron. Astrophys., 30, 543-574, 1992.

Monaghan, J.J., Simulating free surface flows with SPH,
J. Comput. Phys., 110, 399-406, 1994.

Monaghan, J.J., Gravity currents and solitary waves, Physica D, 98,
523-533, 1996.

Monaghan, J.J., and J.C. Lattanzio, A refined particle method for
astrophysical problems,
Astron. Astrophys., 149, 135-143, 1985.

Monaghan, J.J., and R.J. Humble, Vortex particle methods for periodic
channel flow,
J. Comput. Phys., 107, 152-159, 1993.

National Research Council; The role of fluids in crustal processes, Studies in
geophysics, Geophysics study committed, Commission on Geosciences,
Environment and
Ressources, National Academic Press, Washington D.C., 1990.

Ouillon, G., C. Castaing and D. Sornette,
Hierarchical scaling of faulting, J. Geophys. Res., 101, 5477-5487, 1996.

Peitgen, H.-O., and D. Saupe, eds., The Science of Fractal Images,
Springer-Verlag, 1988.

Persson, B.N.J., and E. Tosatti, eds., Physics of sliding friction, NATO
ASI Series
(Kluwer Academic Publishers, Dordrecht, 1996).

Pisarenko, D., and P. Mora, Velocity weakening in a dynamical model of
friction,
Pure and Applied Geophysics, 142, 447-466, 1994.

Rice, J.R., Spatio-temporal complexity of slip on a fault,
J.Geophys.Res., 98, 9885-9907, 1993.

Ruina, A., Slip instability and state variable friction laws,
J. Geophys. Res., 88, 10359-10370, 1983.

Schmittbuhl, J., J.-P. Vilotte and S. Roux,
Velocity weakening friction : A renormalization approach,
J. Geophys. Res., 101, 13911-13917,1996.

Shnirman, M.G.,  and E.M. Blanter, Self-organized criticality in a mixed
hierarchical system,
preprint 1998.

Scholz, C.H., Earthquakes and friction laws, Nature, 391, N6662, 37-42, 1998.

Scott D., Seismicity and stress rotation in a granular model of
the brittle crust, Nature, 381, N6583, 592-595, 1996.

Shaw, B.E., Complexity in a spatially uniform continuum fault model,
Geophys. Res. Lett. 21, 1983-1986, 1994.

Shaw, B.E., Frictional weakening and slip complexity in earthquake faults,
J. Geophys. Res., 102, 18239-18251, 1995.

Shaw, B.E., Model quakes in the two-dimensional wave equation,
J. Geophys. Res., 100, 27367-27377, 1997.

Sokoloff, J.B., Theory of dynamical friction between idealized sliding
surfaces,
Surf. Sci., 144; 267-272, 1984.

Sornette, D., Acoustic waves in random media: III- Experimental situations,
Acustica, 67, 15-25, 1989.

Sornette, D., P. Miltenberger and C. Vanneste, Statistical physics of
fault patterns self-organized by repeated earthquakes : synchronization
versus self-organized
criticality,  in Recent Progresses in Statistical Mechanics and Quantum
Field Theory,
Proceedings of the conference Statistical Mechanics and Quantum Field Theory,
 eds. P. Bouwknegt, P. Fendley, J. Minahan, D. Nemeschansky, K. Pilch,
H. Saleur and N. Warner, World Scientific, Singapore, 1995, pp .313-332.

Stauffer, D., and A. Aharony, Introduction to percolation theory, 2nd ed.,
London ; Bristol, PA : Taylor \& Francis, 1994.

Stellingwerf, R.F. and C.A. Wingate,
Impact modeling with smooth particle hydrodynamics,
In 1992 Hypervelocity Impact Symposium, 1992.

Streit, J.E., Low frictional strength of upper crustal faults: A model,
J. Geophys. Res., 102, 24619-24626, 1997.

Tanguy, A., and P. Nozi\`eres, First-order bifurcation landscape in a 2D
geometry -
The example of solid friction, J. Physique I, 6, 1251-1270, 1996.

Tanguy, A and S. Roux, Memory effects in friction over correlated surfaces,
     Phys. Rev. E, 55, 2166-2173, 1997.

Tillotson, J.H., Metallic Equations of State for Hypervelocity Impact,
General Atomic report GA-3216, 1962.

Tsutsumi, A., and T. Shimamoto, Frictional properties of monzodiorite and
gabbro during
seismogenic fault motion, J. Geol. Soc. Japan, 102, 240-248, 1996.

Tsutsumi, A., and T. Shimamoto, High-velocity frictional properties of gabbro,
Geophys. Res. Lett., 24, 699-702,1997.

\pagebreak

{\bf Figure captions}
\vskip 0.5cm

Fig. 1: Classical friction experiment in which a block of mass $M$ in contact
with a solid substrate is submitted to a normal pressure $P$ and to a
constant horizontal
velocity $v$. We have worked with a block of size $0.5 \times 0.375
\times 0.25~cm^3$ while the substrate has dimension  $1 \times 0.5 \times
0.25~cm^3$.

\vskip 0.5cm

Fig. 2:  A configuration where the particles making up the block and
substrate are represented. In
this example, the total number of particles in the block is $12 \times 16
\times 20 = 1520$.
The size of a particle is of order $0.025~cm$. A slider block of a
centimeter scale
is constituted of effective particles of a fraction of a millimeter
that act as constitutive grains. Each particle at the boundary is
an elementary asperity that interacts with the particle-asperities of the
substrate.

\vskip 0.5cm

Fig. 3: a) Local friction coefficient $\mu \equiv f^x_{spr}/f^z_{spr}$ as
defined in
(\ref{rfchhjkx}) measured on a single particle in the first layer of the
slider block
in contact with the substrate. The simulation uses the boundary force
$f_{1}$ defined by (\ref{eq-f1}). The total number of particles is $5504$, the
sliding velocity is $v = 1~m/s$, the applied pressure is $10^{8}$ Pa, the
dissipation
parameters are $\alpha=0.1, \gamma=10^{6}~s^{-1}$. The time average of this
local
instantaneous friction coefficient is $\mu \approx -0.05$, the
negative sign corresponding to a net drag.

b) Global instantaneous
friction $\mu_l$, obtained by taking the ratio of the total force along $x$
on all block particles on
the boundary in contact with the substrate to the total force along $z$. We
obtain the
same estimate $\mu_l \approx -0.05$ for the friction coefficient when time
averaging.

\vskip 0.5cm

Fig. 4: Vertical motion of a particle of the slider block in the layer
in contact with the substrate, using the boundary force $f_{1}$, $5504$
particles,
$v = 1~m/s$, a pressure of $10^{8}~Pa$, $\alpha=0.1, \gamma=10^{6}~s^{-1}$.

\vskip 0.5cm

Fig. 5: Velocity along $x$ (fig. 5a) and along $z$ (fig.5b)
 of the center of mass of the slider block,
under the same conditions as for figure 4.

\vskip 0.5cm

Fig. 6: Instantaneous friction coefficient
measured on all the particles in the boundary layer in contact with the
substrate
using the boundary force $f_{1}$, $5504$ particles, $v = 10 ~m/s$ which was
imposed at $t = 2~ms$,
a pressure of $10^{8}~Pa$ and dissipation $\alpha=0,1, \gamma=10^{6}~s^{-1}$.
The flat steps in the graph at the value $\mu = 0$ correspond to jumps of
the slider block over the
substrate asperities.

\vskip 0.5cm

Fig. 7: Motion of the slider block over a step of height $h$ after having been
accelerated $v =1~m/s$ before the
step. The simulations use the substrate force $f_{1}$, $5328$ particles
forming cubic lattices
and a normal pressure of $10^6~Pa$. The arrows
represent the instantaneous velocities of the particle.

a) Snapshot exactly
at the time when the slider block encounters the step.

b) The slider block is ejected vertically and starts to jump over the step.

\vskip 0.5cm

Fig. 8: Same as figure 7 but showing a latter time for different parameters
$v = 10~m/s$ and a normal pressure of $10^8~Pa$.
The arrows show the stress carried by each particle
projected in the 2D $(x,z)$ plane.

\vskip 0.5cm

Fig. 9: Measured friction coefficient $\mu_l$ as a function of time for a
simulation using
$f_{1}$, $6640$ particles forming cubic lattices, $v = 1~m/s$ and a normal
pressure of $10^8~Pa$.
The substrate is made of a single layer of particles positionned
regularly on a lattice in the $x-y$ plane, but with their vertical
positions taken randomly and
uniformly between $-\Delta z/2$ and $+\Delta z/2$. The slider block is not
modified.

\vskip 0.5cm

Fig. 10: Two fractal surfaces, respectively with fractal dimensions $D_f =
2.3$ (fig.10a) and
$D_f = 2.8$ (fig.10b), with the same maximum amplitude equal to $h$.
As for fig.9, the substrate is made of a single layer of particles, positionned
regularly on a 2D cubic lattice in the $x-y$ plane with mesh $h$.
Their vertical positions are determined by using the spectral synthesis
method described in
[{\it Peitgen and Saupe}, 1988] to generate a self-affine surface. The
slider block is made of
$6440$ particles organized in a hexagonal compact lattice.
The largest wavelength in the fractal surfaces is
equal to one eighth the length of the slider block to ensure the stability
of the slider
block during its motion.

\vskip 0.5cm

Fig. 11: Locii of contacts between the substrate and the slider block for
the two surfaces
shown in figure 10. Figure 11a, corresponding to $D_f = 2.3$,
 shows a larger and more coherent contact area than figure 11b,
corresponding to $D_f = 2.8$.

\vskip 0.5cm

Fig. 12: Variation of the solid friction coefficient $\mu_l$ as a function of
the substrate fraction dimension $D_f$ for simulations performed under a
normal
pressure of $10^8~Pa$, $v= 1~m/s$ and $\gamma = 10^6~s^{-1}$.

\end{document}